\documentstyle[12pt]{article}
\begin{document}
\newcommand{\LS}{{L^{(S)}}}
\newcommand{\LD}{{L^{(D)}}}
\hoffset 0.5cm
\voffset -0.4cm
\evensidemargin 0.0in
\oddsidemargin 0.0in
\topmargin -0.0in
\textwidth 6.7in
\textheight 8.7in

\begin{titlepage}

\begin{flushright}
PUPT-\\
hep-th/9612223\\
December 1996
\end{flushright}

\vskip 0.2truecm

\begin{center}
{\large {\bf  Four-Brane and Six-Brane Interactions\\
                            in M(atrix) Theory}}
\end{center}

\vskip 0.4cm

\begin{center}
{Gilad Lifschytz}
\vskip 0.2cm
{\it Department of Physics,
     Joseph Henry Laboratories,\\
     Princeton University, \\
     Princeton, NJ 08544, USA.\\ 
    e-mail: Gilad@puhep1.princeton.edu }

\end{center}

\vskip 0.8cm

\noindent {\bf Abstract} 
We discuss the proposed description of configurations with four-branes and six-branes
in m(atrix) theory. Computing the velocity dependent potential 
between these configurations and 
gravitons and membranes, we  show that they agree with the short distance
string results computed in type IIa string theory. Due to the ``closeness'' of 
these configuration to a supersymmetric configuration
the m(atrix) theory reproduces the correct long distance behavior.

\noindent                
\vskip 6 cm

\end{titlepage}
\section{introduction}
Recently \cite{bfss} there has been a proposal for the microscopic description
of M-theory \cite{wit1,ht} in the infinite momentum frame.
In this proposal the only degrees of freedom are the zero-branes and the
lowest open string-modes stretching between them \footnote{ Another approach
can be found in \cite{per}}, giving an $SU(N)$ Yang-Mills theory \cite{wit}.
In order to describe M-theory one has
to recover its brane content \cite{dlp,gr,pol},
  Lorentz invariance, long distance behavior, and correct
compactifications.
Compactification of m(atrix) theory were considered in
\cite{bfss,tay,sus,grt,ks,dg} and  the long distance behavior of membranes
was analyzed in \cite{ab,gilmat}.
The description of the membrane was already given in \cite{bfss}, a description
of an open membrane was given in \cite{li} and a 
proposed description of the four-brane of type IIa (a wrapped five brane of M-theory)
was given in \cite{berdoug}. The four-brane construction however involved introducing
more degrees of freedom into the theory. 

A different approach was suggested in \cite{grt,bss}. 
In \cite{bss} the supersymmetric algebra of zero-brane in the m(atrix) theory was 
analyzed, and it was shown that one has a conserved charge associated with the
membrane and four-brane descriptions. 
They also showed how to construct in this frame work 
configurations with any dimensional brane.

Consider the membrane description in m(atrix) theory. 
The membrane is described through
its effect on the zero-branes bound to it in a non threshold bound state. This can be
seen \cite{grt,gilmat}
by comparing the m(atrix) description to a type IIa description in which one takes
a two-brane with a magnetic field on its world volume. Thus
one can expect to be able to describe any brane in type IIa theory if it can be put in
a non-threshold bound state with zero-branes. In the type IIa description,
 due to the coupling of the two-brane to $RR$ background
 ($A$ is the $RR$ one-form gauge potential)
$\int A \wedge {\cal F} $ \cite{doug},
the zero-branes are taken into account by turning on a magnetic field on the
two-brane. In order to bound a four-brane in 
 a non-threshold bound state with zero-branes one can use the four brane coupling to
$A_{\mu}$, $\frac{1}{2}\int A \wedge {\cal F} \wedge {\cal F}$, with a constant magnetic
field {\cal F}. This  also adds two-branes, 
through the coupling $\int C \wedge {\cal F}$, where $C$ is the $RR$ three-form
gauge potential. 

In the two-brane case the matrix description \cite{bfss} was achieved by
taking $[X_1, X_2]=Iic$, it is then natural to take for the four-brane, 
four matrices which satisfy $[X_1,X_2]=Iic_1$ and $[X_3,X_4]=Iic_2$ \cite{bss}.
This can be  generalized to higher dimensional branes.
The Four-brane constructed this way will also have membranes (and off course
zero-branes) bound to it, and the six-branes will have four-branes and membranes
bound to it. We will however in this paper call them a four-brane and 
a six-brane.

In this paper we analyze this construction. we compute the potential between 
configurations involving six-branes, four-branes, two-branes and zero-branes.
The potentials are compared with calculations in the type IIa theory of the 
corresponding configurations. In all cases (as in \cite{gilmat})
we find exact agreements between the
m(atrix) description and the type IIa short distance description.
Due to the ``closeness'' of the type IIa configuration to being supesymmetric, we
find that the short distance and long distance potentials agree \cite{dkps}, thus
enabling the m(atrix) description to describe long distance potentials as well.

It should be noted that the description studied in this paper is only one
out of possible constructions for configurations involving four-branes and six-branes.

\section{The Calculation}
In this section we will calculate the potential between various configurations
of gravitons, membranes, four-branes and six-branes in M(atrix) theory.
In section (2.1) we describe the interaction
between a four-brane and a zero-brane,  in section (2.2) the interaction between a 
four-brane and a membrane parallel to it,  in section (2.3) the interaction between
two parallel four-branes and in section (2.4) we 
describe the interaction between a six-brane
and a zero-brane.

Let us start with the Lagrangian \cite{wit,bfss,kp,dfs}, we take the string
length $l_{s}=1$, the signature is $(-1,1 \ldots,1)$, and  
$D_t X=\partial_t X -i [A_0,X]$,
\begin{equation}
L=\frac{1}{2g}Tr\left[D_{t}X_{i}D_{t}X^{i}+2\theta^{T}D_{t}\theta-
\frac{1}{2}[ X^{i},X^{j}]^{2}-2\theta^{T}\gamma_{i}[\theta,X^{i}] \right].
\label{l}
\end{equation}
The supersymmetry transformations are
\begin{eqnarray}
\delta X^{i}=-2\epsilon^{T} \gamma^{i} \theta \nonumber\\
\delta \theta =\frac{1}{2}\left[ D_{t}X^{i}\gamma_{i}+\frac{1}{2}
[X^i, X^j]\gamma_{ij} \right]\epsilon +\epsilon' \nonumber \\
\delta A_{0} =-2\epsilon^{T}\theta
\label{susy}
\end{eqnarray}
We chose to work in the background covariant gauge (the ghost will be
called $C$).  We give certain 
$X$'s some expectation value $B$ and  write $X_i=B_i +Y_i$.
If one chooses the $B_{\mu}$ such that $B_0=0$ and the other $B_{i}$ solve the
equation of motion then we can expand the Lagrangian 
to quadratic order in the fluctuations
around the background fields and find 
\begin{eqnarray}
L_{2} & = & \frac{1}{2g}Tr \{ (\partial_0 Y_{i})^2-(\partial_0 A_{0})^2
-4i\dot{B}_i[A_0,Y^i]+\frac{1}{2}[B_i,Y_j]^2+\frac{1}{2}[B_j,Y_i]^2  \\
  & + & [B_i,Y^j][Y^i,B^j]+[B_i,Y^i][B_j,Y^j]-
[A_0,B_i]^2+[B_i,B_j][Y^i,Y^j]  \nonumber \\
 & + & \partial_{0} C^* \partial_{0} C+[C^*,B^i][B_i,C] 
+2\theta^{T} \partial_{t} \theta -2 \theta^{T} \gamma_{i}[\theta,B^i]\}.
\label{l3}
\end{eqnarray}

We take the following form for $Y_i$ and $\theta$.
\[
Y_i=\left(
\begin{array}{cc}
0 & \phi_i \\ 
\varphi_i & 0 
\end{array}
\right), \ \
\theta=\left(
\begin{array}{cc}
0 & \psi \\ 
\chi & 0 
\end{array}
\right)
\]
Where in matrix space $\varphi = \phi ^{\dagger}$ and $\chi = \psi^{T}$.

\subsection{Four-brane zero-brane scattering}
The background configuration for a zero-brane scattering off a
four-brane is \cite{bss},
\[
B_8=\left(
\begin{array}{cc}
P_1 &  0  \\ 
0 & 0 
\end{array}
\right),
B_9=\left(
\begin{array}{cc}
Q_1 &  0  \\ 
0 & 0 
\end{array}
\right),
B_7=\left(
\begin{array}{cc}
P_2 &  0  \\ 
0 & 0 
\end{array}
\right),
B_6=\left(
\begin{array}{cc}
Q_2 &  0  \\ 
0 & 0 
\end{array}
\right),
\]
\[
B_5=\left(
\begin{array}{cc}
bI &  0  \\ 
0 & 0 
\end{array}
\right),
B_1=\left(
\begin{array}{cc}
Ivt &  0  \\ 
0 & 0 
\end{array}
\right).
\]
where $[Q_1,P_1]=ic_1$ , $[Q_2,P_2]=ic_2$ and we will soon discuss what are the
values of $c_1,c_2$. The four-brane is thought of as wrapped on a large
$T^4$ of radiuses $(R_9,R_8,R_7,R_6)$ respectively.

The graviton scattering off the four-brane is given to leading order by multiplying
the result for the zero-brane, by the number of zero-branes the graviton is made of.

In order to calculate the potential we should compute the mass matrix for $\phi$ and
$\psi$, and then compute the one loop vacuum energy by evaluating the 
determinant of the operator ${\det}(\partial_{t}^{2} + M^2)$.
Now if we think of $\phi$ and $\psi$ as $N$ dimensional vectors (i.e the total number
of zero-brane in this bound state $N$) then we should understand the $P_1,Q_1$
and $P_2, Q_2$ matrices as only  $N_1 \times N_1$ and  $N_2 \times N_2$ 
matrices, with $N_1 N_2=N$,  as explained
in \cite{bss}.  We will shortly
see what this means in term of number of two-brane and zero-branes
bounded on the four-brane.

Inserting the above background into equation (\ref{l3}), we find that the mass
matrix squared, in the space of $(Y_2 \ldots Y_5,C)$ is proportional to the identity with
the proportionality constant being $2H$, and 
\begin{equation}
H=P_{1}^{2}+P_{2}^{2}+Q_{1}^{2}+Q_{2}^{2}+Iv^2 t^2 +I b^2.
\end{equation}

In the space of $A_0,Y_1$ there are also off diagonal terms of $\pm 4iv$
\[
M^2_{A_0Y_1}=2\left(
\begin{array}{cc}
-H&  -2i v \\ 
2i v & H
\end{array}
\right)
\]
In the space of $Y_8,Y_9$ one has also off diagonal terms $\pm4ic_1$, and
in the space of $Y_7,Y_6$ one has off diagonal terms $\pm 4ic_2$, thus
\[
M^2_{Y_8Y_9}=2\left(
\begin{array}{cc}
H&  -2 ic_1\\ 
2 ic_1 & H
\end{array}
\right),
M^2_{Y_7Y_6}=2\left(
\begin{array}{cc}
H&  -2 ic_2\\ 
2 ic_2 & H
\end{array}
\right).
\]
 Evaluating the fermionic terms we find 
\begin{equation}
m_{f}=\gamma_8 P_1 +\gamma_9 Q_1 +\gamma_7 P_2 + \gamma_6 Q_2
+ \gamma_1 Ivt +\gamma_5 Ib
\end{equation}
We now rotate to Euclidean space ($t=i\tau$ , $A_{0}=-iA_{\tau}$), and
convert the fermionic determinants to a form $(\det(-\partial^{2}_{\tau} +M_{f}^2))$.
\begin{equation}
M^{2}_{f}=H+Iic_1 \gamma_9 \gamma_8 +Iic_2 \gamma_6 \gamma_7 
+Iv \gamma_1.
\end{equation}
This gives for the bosonic determinants two (complex) bosons with $M^2=2H$, one
with $M^2=2H+4iv$, one with $M^2=2H-4iv$, one with
$M^2=2H+4c_1$, one with $M^2=2H-4c_1$, one with $M^2=2H+4c_2$
and one with $M^2=2H-4c_2$. From the fermionic fields we get 
determinants with $M^{2}_{f}$. Two with $M^{2}_{f}=H+c_1+c_2+iv$, two
with $M^{2}_{f}=H+c_1+c_2-iv$, two with $M^{2}_{f}=H-c_1-c_2+iv$, two
with $M^{2}_{f}=H-c_1-c_2-iv$, two with $M^{2}_{f}=H-c_1+c_2+iv$, two
with $M^{2}_{f}=H-c_1+c_2-iv$, two with $M^{2}_{f}=H+c_1-c_2+iv$ and  two
with $M^{2}_{f}=H+c_1-c_2-iv$.

How do the $P$'s and $Q$'s act on $\phi$ and $\psi$ ?. One can realize these operator
on the space of functions of two variables $(x,y)$.
Then $P_1$ can be realized as $-ic_1 \partial_{x}$,  $Q_1$ as
$x$, $P_2$ as   $-ic_2 \partial_{y}$ and $Q_2$ as $y$. 
The spectrum of $H$ is then, 
\begin{equation}
H_n = b^2 +v^2 t^2 +c_1(2n_1+1) +c_2(2n_2+1)
\end{equation}
Define $r^{2}_{n_1 n_2}=b^2 +c_1(2n_1+1) +c_2(2n_2+1)$ then
the phase shift of a zero-brane scattered off the four-brane
configuration, to one-loop is
\begin{eqnarray}
\delta & = &\frac{1}{2}\sum_{n_1n_2} \int \frac{ds}{s} e^{-sr^{2}_{n_1 n_2}}
\frac{1}{\sin sv} [2+2\cos 2vs +2\cosh 2sc_1 +2\cosh 2sc_2  \nonumber\\
 & - & 4\cos vs (\cosh (c_1 + c_2)s +\cosh (c_1 -c_2)s )].
\end{eqnarray}
Summing over $n_1, n_2$ we get
\begin{equation}
\delta=\int \frac{ds}{s} e^{-sb^2}
\frac{2+2\cos 2vs +2\cosh 2sc_1 +2\cosh 2sc_2  
-4\cos vs (\cosh (c_1 + c_2)s +\cosh (c_1 -c_2)s )}{8\sinh sc_1 \sinh sc_2 \sin sv}.
\label{m40}
\end{equation}
Notice there is a tachyonic instability \cite{bansus} for $b^2 < |c_1 - c_2|$.

let us compare this with the corresponding string configuration of a four-brane
with many two-branes (orthogonally embedded) 
and many zero-branes, all bounded in a non-threshold 
bound state.
The phase shift of a zero-brane scattering off this bound state was computed in
\cite{gil2}, where it was called the $(4-2-2-0)$ bound state (for the classical
supergravity solution see \cite{bmm}, and for a T-dual description see \cite{bdl}). 
The string configuration is described by a four-brane with a world-volume 
magnetic field turned on (some configurations were discussed in \cite{bacpor,gregut}).
,

\[
{\cal F}=\left(
\begin{array}{ccccc}
0 & 0 & 0 & 0 & 0 \\
0 & 0 & F_1 & 0 & 0 \\
0 & -F_1& 0 & 0 & 0 \\
0 & 0 & 0 & 0 & F_2 \\
0 & 0 & 0 & -F_2 & 0
\end{array}
\right).
\]
This describes a four-brane with two-branes in the $8,9$ direction,
two-branes in the $6,7$ direction and some zero-branes.
We choose to take $F_1$ in the $8,9$ direction and $F_2$ in the $6,7$ direction.
Notice that $F_1$ describes a two-brane in the four-brane stretched in the
$6,7$ direction $F_2$ a two-brane in the $8,9$ direction.
From the coupling of a D-brane to a $RR$ background \cite{doug}
one can read off the number of
embedded two-branes (call them $n_1$ and $n_2$). $2\pi R_8 R_9 F_1= n_1$,
$2\pi R_6R_7 F_2= n_2$, and the number of zero-branes 
$N=n_1 n_2$ 

Define $\tan \pi \epsilon_{j} =F_{j}$. Using the same notation as in 
\cite{gil2}, and $\Theta(\rho)=\Theta(\rho,is)$,  the phase shift takes the form,
\begin{equation}
\delta_{IIA}=\frac{1}{2\pi}\int \frac{ds}{s} e^{-b^2 s}
 B \times J.
\label{ampf4220}
\end{equation}
\begin{eqnarray}
B& = &\frac{1}{2}f_{1}^{-4}\Theta^{-1}_{4}(i\epsilon_1 s)\Theta^{-1}_{4}(i\epsilon_2 s)
\frac{\Theta_{1}' (0)}{\Theta_{1}(\nu t)}. \nonumber\\
J& = &\{ -f_{2}^{4}
\frac{\Theta_{2} (\nu s)}{\Theta_{2}(0)}
\Theta_{3}(i\epsilon_1 s)\Theta_{3}(i\epsilon_2 s)
+f_{3}^{4}\Theta_{2}(i\epsilon_1 s)\Theta_{2}(i\epsilon_2 s)
\frac{\Theta_{3} (\nu s)}{\Theta_{3}(0)} \nonumber \\
 & + & f_{4}^{4} 
\frac{\Theta_{4} (\nu s)}{\Theta_{4}(0)}\Theta_{1}(i\epsilon_1 s)
\Theta_{1}(i\epsilon_2 s)\}.
\label{f40}
\end{eqnarray}

If $F$ is very large, 
let $\epsilon_j=\frac{1}{2}-c_{j}'$ ($c'$ is very small), 
the phase shift  becomes ($\tanh \pi \nu =v$)
\begin{equation}
\delta_{IIA}=\frac{1}{2\pi}\int \frac{ds}{s} e^{-b^2 s}
 B \times J.
\label{amp4220}
\end{equation}
\begin{eqnarray}
B& = &-\frac{1}{2}f_{1}^{-4}\Theta^{-1}_{1}(ic_{1}'s)\Theta^{-1}_{1}(ic_{2}'s)
\frac{\Theta_{1}' (0)}{\Theta_{1}(\nu t)}. \nonumber\\
J& = &\{ -f_{2}^{4}
\frac{\Theta_{2} (\nu s)}{\Theta_{2}(0)}
\Theta_{2}(ic_{1}'s)\Theta_{2}(ic_{2}'s)+f_{3}^{4}
\Theta_{3}(ic_{1}' s)\Theta_{3}(ic_{2}' s)
\frac{\Theta_{3} (\nu s)}{\Theta_{3}(0)} \nonumber \\
 & - & f_{4}^{4} 
\frac{\Theta_{4} (\nu s)}{\Theta_{4}(0)}\Theta_{4}(ic_{1}'s)\Theta_{4}(ic_{2}'s)\}.
\label{c40}
\end{eqnarray}

First let us evaluate (\ref{c40}) as if only the massless open string mode
would have contributed (i.e very short distances).  We find ($\pi c' =c$)
\begin{equation}
B \times J =\pi
\frac{2+2\cos 2vs +2\cosh 2sc_1 +2\cosh 2sc_2  
-8\cos vs (\cosh sc_1 \cosh sc_2)}
{4\sinh sc_1 \sinh sc_2 \sin sv}
\label{s40}
\end{equation}
In this limit we get exactly the result from the M(atrix) approach.
This is another example that in some sense the M(atrix) description is just another
description of a type IIA calculation.

Now we can evaluate $c$. From the definition of $c'$ one finds
$F_{j}=c^{-1}_{j}$, we already know the relationship between $F$ and
the number of branes, then
$c_1=\frac{2\pi R_8 R_9}{n_1}$ and $c_2=\frac{2\pi R_6 R_7}{n_2}$.
$n_1$ and $n_2$ are then the number of zero-branes on the bounded 
two-branes in the $8,9$ and $6,7$ directions respectively. Identifying 
$N_1=n_1$ and $N_2=n_2$, this explains 
why the $P$'s and $Q$'s were $N_1 \times N_1$ and $N_2 \times N_2$
dimensional matrices.  This is consistent with having 
$N_1$ and $N_2$ two-branes in the $6,7$ and $8,9$ directions
respectively and with having a total of $N$ zero-branes in the bound state.

We can now calculate the long range potential from the M(atrix) calculation and
from the string calculation (keeping now the lowest modes of the closed string).
Both calculations agree to lowest order in $c$ and $v$ and we find

\begin{equation}
V=-\Gamma(3/2)\frac{2v^2 (c_{1}^{2}+ c_{2}^{2}) + v^4 + (c_{1}^{2} -c_{2}^{2})^2}
{8\sqrt{\pi} c_1 c_2 } b^{-3}.
\label{v40}
\end{equation}
From equation (\ref{v40}) we see that if $c_1 =c_2$, then there is 
no force if there is no relative velocity. This is because this configuration then
 preserves a quarter of the supersymmetry \cite{bdl,gil2}, and  is a signature
of the presence of the four-brane.
The agreement of the long distance potentials shows that the four-brane constructed 
this way has the right tension.


\subsection{Four-brane membrane interaction}
In this subsection we will compute the velocity dependent potential between a 
four-brane and a membrane parallel to each other in the m(atrix) theory.
The background configuration is
\[
B_8=\left(
\begin{array}{cc}
P_1 &  0  \\ 
0 & P_3 
\end{array}
\right),
B_9=\left(
\begin{array}{cc}
Q_1 &  0  \\ 
0 &  Q_3
\end{array}
\right),
B_7=\left(
\begin{array}{cc}
P_2 &  0  \\ 
0 &  0
\end{array}
\right),
B_6=\left(
\begin{array}{cc}
Q_2 &  0  \\ 
0 &  0
\end{array}
\right),
\]
\[
B_5=\left(
\begin{array}{cc}
bI &  0  \\ 
0 & 0 
\end{array}
\right),
B_1=\left(
\begin{array}{cc}
Ivt &  0  \\ 
0 & 0 
\end{array}
\right),
\]
and $c_1=c_2=c_3=c$.
Inserting this background to equation (\ref{l3}) we find the mass matrix for the
bosons and fermions.
Define
\begin{equation}
H=(P_1 +P_3)^2 +(Q_1- Q_3)^2 +P_{2}^{2}+Q_{2}^{2}+Ib^2 +Iv^2 t^2.
\end{equation}
The mass matrix squared for the bosons in the space $(Y_2 \ldots Y_5,Y_8,Y_9,C)$
is $2IH$. For the other bosons we find
\[
M^2_{A_0Y_1}=2\left(
\begin{array}{cc}
-H&  -2i v \\ 
2i v & H
\end{array}
\right),
M^2_{Y_6Y_7}=2\left(
\begin{array}{cc}
H&  -2 ic_2\\ 
2 ic_2 & H
\end{array}
\right).
\]
For the fermions one finds
\begin{equation}
m_{f}=\gamma_8 (P_1 + P_3) +\gamma_9 (Q_1 -Q_3) + \gamma_7 P_2
+\gamma_6 Q_2 +\gamma_1 Ivt + \gamma_5 Ib.
\end{equation}

 After rotating to Euclidean space and converting the fermion determinant to the form
$\det(-\partial^{2}_{\tau} + M_{f}^{2})$ we find the following:
four complex bosons with $M^2= 2H$, one with $M^2= 2H+4iv$, one with
$M^2= 2H-4iv$, one with $M^2= 2H+4c$ and one with $M^2= 2H-4c$.  
For the fermions there is four with $M_{f}^{2}=H+c+iv$, four with
$M_{f}^{2}=H+c-iv$, four with $M_{f}^{2}=H-c+iv$ and four with
$M_{f}^{2}=H-c-iv$.
The $P$'s and $Q$'s can be represented  as $Q_1-Q_3=x_1$, $Q_1+Q_3=y_1$,
$Q_2=x_2$, $P_1 + P_3=2ic\partial_{y_1}$, $P_2=2ic\partial_{x_2}$ and
$P_1 -P_3=2ic\partial_{ x_1}$.
The spectrum of $H$ is then
\begin{equation}
H_{n,x_1,k_1}=b^2 +v^2 t^2 +c(2n+1)+4c^2 k_{1}{^2} +x^{2}_{1},
\end{equation}
and $H$ has a degeneracy which we label by $N_{-}$ \cite{gilmat}.
Evaluating the determinants, summing over $n$ 
and integrating over $(x_1,k_{1})$, the phase shift is
\begin{equation}
\delta=N_{-}\int \frac{ds}{s} e^{-b^2s}\frac{4+2\cosh 2cs +2\cos 2vs -8\cos vs \cosh cs}
{16cs \cosh cs}.
\label{m42}
\end{equation}

The string description is that of a four-brane with a magnetic field on its world volume
(as in section (2.1)), and a two-brane with a magnetic field on its world
volume as in \cite{gilmat}. As we took all the $c$'s to be equal we should take all 
the magnetic fields to be equal and large. 
The phase shift for the above two-brane when
scattered off the above four-brane configuration 
($\tan \pi (1/2 -c')=F$, $\tanh \pi \nu =v$) 
\begin{equation}
\delta_{IIa}=\frac{L^2 (1+F^2)}{2\pi} \int \frac{ds}{s}\frac{e^{-b^2s}}{4\pi s}
B \times J.
\label{d42}
\end{equation}
Where $L^2$ is the volume of the two-brane and
$B\times J$ is the same as in the case of a zero-brane scattering off a two-brane 
with a magnetic field on its world volume \cite{gilmat} 
\begin{eqnarray}
B& = &\frac{1}{2}f_{1}^{-6}(-i\Theta_{1})^{-1}(ic's)
\frac{\Theta_{1}' (0)}{\Theta_{1}(\nu s)}, \nonumber \\
J& = &\{ -f_{2}^{6}
\frac{\Theta_{2} (\nu s)}{\Theta_{2}(0)}
\Theta_{2}(ic` s)+f_{3}^{6}\Theta_{3}(ic' s)
\frac{\Theta_{3} (\nu s)}{\Theta_{3}(0)} \nonumber\\ 
 & - & f_{4}^{6} 
\frac{\Theta_{4} (\nu s)}{\Theta_{4}(0)}\Theta_{4}(ic' s)\}.
\label{bj42}
\end{eqnarray}
If we now evaluate equation (\ref{bj42}) in the limit that the only contribution 
comes from the lowest modes of the open string, and insert that into equation
(\ref{d42}) we find that $\delta_{IIa}=\delta$ (when one identifies $\pi c' =c$, and
$N_{-}=\frac{L^2}{\pi c}$ as in \cite{gilmat}). 
Thus the m(atrix) calculation
agrees with the short distance string calculation.

Comparing the long distance potentials from the string theory and from the m(atrix)
theory we find that to leading order in $v$ and $c$ they agree and give
\begin{equation}
V=-\frac{L^2\Gamma(3/2)(2v^2 c^2 +c^4 +v^4)}{16\pi^{5/2} c^3} b^{-3}.
\end{equation}

\subsection{Four-brane four-brane interaction}
In this subsection we will consider the interactions of two of the above
four-brane in M(atrix) theory.
The background configuration is 
\[
B_8=\left(
\begin{array}{cc}
P_1 &  0  \\ 
0 & P_3 
\end{array}
\right),
B_9=\left(
\begin{array}{cc}
Q_1 &  0  \\ 
0 &  Q_3
\end{array}
\right),
B_7=\left(
\begin{array}{cc}
P_2 &  0  \\ 
0 &  P_4
\end{array}
\right),
B_6=\left(
\begin{array}{cc}
Q_2 &  0  \\ 
0 &  Q_4
\end{array}
\right),
\]
\[
B_5=\left(
\begin{array}{cc}
bI &  0  \\ 
0 & 0 
\end{array}
\right),
B_1=\left(
\begin{array}{cc}
Ivt &  0  \\ 
0 & 0 
\end{array}
\right).
\]
Where we take $c_1=c_2=c_3=c_4=c$.
Define 
\begin{equation}
H=(P_1 + P_3)^2 +(Q_1 - Q_3)^2 +(P_2 + P_4)^2 +(Q_2 - Q_4)^2 
+b^2I + I v^2 t^2.
\end{equation}
 Inserting the background to equation (\ref{l3}) and computing the mass matrix 
(in Euclidean space), we find
for the complex bosons: six with $M^2=2H$ one with $M^2=2H+4iv$ and
one with $M^2=2H-4iv$. 
For the fermions: eight with $M_{f}^{2}=H+iv$ and eight with
$M_{f}^{2}=H-iv$.

The spectrum of $H$ is continues and there is a degeneracy as in the case of
two membranes \cite{gilmat}. We can realize  $Q_1-Q_3=x_1$, 
 $Q_2-Q_4=x_2$,  $Q_1+Q_3=y_1$ and  $Q_2+Q_4=y_2$. Then
$P_1+P_3 =-2ic\partial_{y_1}$ , $P_1-P_3 =-2ic\partial_{x_1}$  and 
similarly for $P_2,P_4$. The degeneracy will be labeled
by $N_{-}$.

The phase shift is then
\begin{equation}
\delta=8N_{-}^{2}\int_{-\infty}^{\infty} dx_1 dx_2 \frac{dk_1}{2\pi}
\frac{dk_2}{2\pi}\int \frac{ds}{s} e^{-sr^2_{(x_i,k_i)}}
\frac{\sin^{4} (sv/2)}{sin sv}
\end{equation}
where 
\begin{equation}
r^2_{(x_i,k_i)}=b^2+4c^2(k_{1}^{2}+k_{2}^{2})+x_{1}^{2}+x_{2}^{2}.
\end{equation}
Doing the integrals and evaluating the potential one finds
\begin{equation}
V=-\frac{N_{-}^{2} \Gamma(3/2) v^4}{32c^2 \sqrt{\pi} b^{3}}.
\label{44m}
\end{equation}

The long range string calculation using  \cite{bac,gil1,bacpor,gregut} gives,
\begin{equation}
V_{IIA}=-\Gamma(3/2)\frac{(1+F^2)^2 L^4 v^4}{32 \pi^{5/2} b^3}
\label{s44}
\end{equation}
Where $L^4 $ is the area  of $T^4$ on which the four-branes are wrapped.
Using $F=c^{-1}$ and from \cite{gilmat} $N_{-}=\frac{L^2}{\pi c}$ we see
that the string and M(atrix) calculations  agree.

If we would not have taken $c_1 =c_3$ and $c_2= c_4$  we would have gotten a
non zero-force even at $v=0$. One can also make an anti-four-brane by flipping a sign 
of one of the $P$'s or $Q$'s.

\subsection{Six-brane zero-brane scattering}
A six-brane has no bound states with zero-branes. This is because the long
range potential is repulsive $\sim \frac{1}{r}$ and the short distance is
repulsive $\sim r$. The matrix-theory however describes everything in terms
of zero-branes, so one needs to find a configuration with a six-brane that can bind
to zero-branes. This can be achieved by adding four-branes and two-branes
bounded to the six-brane. 
In the same spirit as for the four-brane the configuration of a background of a six-brane 
is

\[
B_8=\left(
\begin{array}{cc}
P_1 &  0  \\ 
0 & 0 
\end{array}
\right),
B_9=\left(
\begin{array}{cc}
Q_1 &  0  \\ 
0 & 0 
\end{array}
\right),
B_7=\left(
\begin{array}{cc}
P_2 &  0  \\ 
0 & 0 
\end{array}
\right),
B_6=\left(
\begin{array}{cc}
Q_2 &  0  \\ 
0 & 0 
\end{array}
\right),
\]
\[
B_5=\left(
\begin{array}{cc}
P_3 &  0  \\ 
0 & 0 
\end{array}
\right),
B_4=\left(
\begin{array}{cc}
Q_3 &  0  \\ 
0 & 0 
\end{array}
\right),
B_3=\left(
\begin{array}{cc}
bI &  0  \\ 
0 & 0 
\end{array}
\right),
B_1=\left(
\begin{array}{cc}
Ivt &  0  \\ 
0 & 0 
\end{array}
\right).
\]
Here we are going to take $c_1=c_2=c_3=c$, and the six-brane is wrapped on
a large $T^6$ with equal sides of length $2\pi R$. As in the case of the 
four-brane, if the total number of zero-branes is $N$ then the 
$P,Q$ matrices should be thought of as $N^{1/3} \times N^{1/3}$ matrices.
One substitutes this background into equation (\ref{l3}), and computes
the mass matrix.
Define 
\begin{equation}
H=P_{1}^{2}+P_{2}^{2}+P_{3}^{2}+Q_{1}^{2}+Q_{2}^{2}+Q_{3}^{2}
+Ib^2 +I v^2 t^2.
\end{equation}
In the space of $(Y_2,Y_3,C)$ $M^2=2IH$, eventually this sector will not
contribute as the ghost will cancel the $Y_2,Y_3$ contributions.
We also find
\[
M^2_{A_0Y_1}=2\left(
\begin{array}{cc}
-H&  -2i v \\ 
2i v & H
\end{array}
\right)
M^2_{Y_8Y_9}=2\left(
\begin{array}{cc}
H&  -2 ic\\ 
2 ic & H
\end{array}
\right).
\]
\[
M^2_{Y_7Y_6}=2\left(
\begin{array}{cc}
H&  -2 ic\\ 
2 ic& H
\end{array}
\right).
M^2_{Y_5Y_4}=2\left(
\begin{array}{cc}
H&  -2 ic\\ 
2 ic& H
\end{array}
\right).
\]
For the fermions we find 
\begin{equation}
M^{2}_{f}=H+vI\gamma_1+ic(\gamma_9 \gamma_8 +\gamma_6 \gamma_7 +
\gamma_4 \gamma_5).
\end{equation}
This gives for the complex bosons (in Euclidean space): one with
$M^2=2H+4iv$, one with $M^2=2H-4iv$, three with $M^2=2H+4c$
and three with $M^2=2H-4c$. 

For the fermions: One with $M_{f}^{2}=H+3c-iv$, one with $M_{f}^{2}=H+3c+iv$,
one with $M_{f}^{2}=H-3c-iv$, one with $M_{f}^{2}=H-3c+iv$, three with
$M_{f}^{2}=H+c-iv$, three with $M_{f}^{2}=H+c+iv$, three with $M_{f}^{2}=H-c-iv$
and three with $M_{f}^{2}=H-c+iv$.

The $P,Q$ matrices are then realized on the space of functions of three variables 
$(x,y,z)$.
The spectrum of $H$ (similarly to the four-brane case)  is
\begin{equation}
H_n=b^2+v^2 t^2 +c(2n_1 +2n_2 +2n_3 +3)
\end{equation}
The phase shift of a scattered zero-brane off this six-brane is then,
\begin{equation}
\delta =\int \frac{ds}{s} e^{-b^2 s} \frac{2\cos 2vs +6\cos 2cs -
\cos vs (2\cosh 3cs +6\cosh cs)}{16\sinh^{3} cs \sin vs}
\label{m60}
\end{equation}
Notice that in this case there is no tachyonic instability. This would not be the case 
if $c_1 \neq c_2 \neq c_3$. 

We turn know to the corresponding string calculation which is a six-brane with a 
world volume magnetic field turn on

\[
{\cal F}=\left(
\begin{array}{ccccccc}
0 & 0 & 0 & 0 & 0 & 0 & 0\\
0 & 0 & F_1 & 0 & 0 & 0 & 0\\
0 & -F_1& 0 & 0 & 0 & 0 & 0\\
0 & 0 & 0 & 0 & F_2 & 0 & 0\\
0 & 0 & 0 & -F_2 & 0 & 0 & 0\\
0 & 0 & 0 & 0 & 0 & 0 & F_3\\
0 & 0 & 0 & 0 & 0 & -F_3 &  0
\end{array}
\right)
\]
In the M(atrix) configuration we took $c_1=c_2=c_3$ so here we take 
$F_1=F_2=F_3=F$. Define $\tan \pi \epsilon = F$.
This configuration describes a six-brane bound with four-branes , two-branes and 
zero-branes.
The phase shift of a zero-brane scattering off this configuration 
is given at  one-loop by,

\begin{equation}
\delta_{IIA}=\frac{1}{2\pi}\int \frac{ds}{s} e^{-b^2 s}
 B \times J.
\label{ampf6420}
\end{equation}
\begin{eqnarray}
B& = &\frac{1}{2}f_{1}^{-2}\Theta^{-3}_{4}(i \epsilon  s)
\frac{\Theta_{1}' (0)}{\Theta_{1}(\nu t)}. \nonumber\\
J& = &\{ -f_{2}^{2}
\frac{\Theta_{2} (\nu s)}{\Theta_{2}(0)}
\Theta^{3}_{3}(i \epsilon s)
+f_{3}^{2}\Theta^{3}_{2}(i \epsilon s)
\frac{\Theta_{3} (\nu s)}{\Theta_{3}(0)} \nonumber \\
 & - &i f_{4}^{2} 
\frac{\Theta_{4} (\nu s)}{\Theta_{4}(0)}\Theta^{3}_{1}(i \epsilon s)\}.
\label{f60}
\end{eqnarray}

If $F$ becomes large it is convenient to define $\epsilon=\frac{1}{2}-c'$, then
the phase shift becomes
\begin{equation}
\delta_{IIA}=\frac{1}{2\pi}\int \frac{ds}{s} e^{-b^2 s}
 B \times J.
\label{amp6420}
\end{equation}
\begin{eqnarray}
B& = &\frac{1}{2}f_{1}^{-2}(-i\Theta_{1}(i c' s))^{-3}
\frac{\Theta_{1}' (0)}{\Theta_{1}(\nu t)}. \nonumber \\
J& = &\{ -f_{2}^{2}
\frac{\Theta_{2} (\nu s)}{\Theta_{2}(0)}
\Theta^{3}_{2}(i c' s)
+f_{3}^{2}\Theta^{3}_{3}(i c' s)
\frac{\Theta_{3} (\nu s)}{\Theta_{3}(0)} \nonumber \\
 & - & f_{4}^{2} 
\frac{\Theta_{4} (\nu s)}{\Theta_{4}(0)}\Theta^{3}_{4}(i c' s)\}.
\label{c60}
\end{eqnarray}
 
We now follow the same route as in the previous subsection. Expanding
equation (\ref{c60}) in the limit when only the lightest open string modes contribute
we find ($\pi c' =c$)
\begin{equation}
B \times J =\pi \frac{6\cosh 2cs + 2\cos 2vs -8\cos vs \cosh^{3} cs}
{8\sinh^{3} cs \sin vs}.
\label{s60}
\end{equation}
Inserting this to the expression for the phase shift and comparing with equation
(\ref{m60}) we find that both expressions are the same.

Now $F=\frac{1}{c}$ and $2\pi R^2=n_4$ the number of four-brane in each direction.
Given there are a total of $N$ zero-brane $n_4=N^{1/3}$, so
$c=\frac{2\pi R^2}{N^{1/3}}$. The number of two-brane in each direction is $N^{2/3}$.

The long range potential from the string calculation can be now compared with the long
range calculation in the M(atrix) theory. To lowest order in $c$ and $v$ they agree
and we find
\begin{equation}
V=-\Gamma(1/2)\frac{v^4-3c^4 +6v^2 c^2}{16 \sqrt{\pi} c^3}b^{-1}.
\end{equation}
the repulsive force coming from the term $\sim c^4$ is due to the six-brane.
Again the agreement of the long distance potentials shows that the six
brane has the right tension.

\section{Conclusions}
In this paper we explored the construction of four-branes and six-branes in
the context of m(atrix) theory. 
We have computed the potential between membranes and zero-branes,
 and configurations in m(atrix)
theory that include four-branes and six-branes. 
These results were shown to be identical to a short distance 
string theory computation in type
IIa, of the corresponding configurations. Due to the large number of bounded zero-branes
on the six-brane and four-brane, these configuration are very close to being 
supersymmmetric \cite{gilmat}. Thus the long distance potentials can be reproduced
by a short distance calculation involving only the lightest open string modes,
so the m(atrix) theory can reproduce the long distance potentials. The agreement of
these calculations supports the proposed description of the four-brane 
and six-brane configurations.

This construction does not give the pure four-brane and six-brane but
rather needs more branes to be added in each case as to achieve a state that
can bind in a non-threshold bound state with zero-branes. Notice that
although the four-brane does have a threshold bound state with zero-branes
without the addition of two-branes, the six-brane has no bound states with 
zero-branes without extra branes added. So while we may hope to be 
able to describe a pure four-brane, there  seem to be an obstacle 
to describe a pure six-brane.
 
\begin{center}
{\bf Acknowledgments}
\end{center}
I would like to thank S.D. Mathur for helpful discussions.

\end{document}